\documentclass{article}
\usepackage{graphics}

\renewcommand{\slash}[1]{#1\llap{/}}
\newcommand{\Tr}{\mathop{\mathrm{Tr}}\nolimits}

\begin{document}

\begin{flushright}
TTP 97--04\\
Budker INP 97--2\\
hep-ph/9701415
\end{flushright}
\vspace{5mm}
\begin{center}
{\huge HQET chromomagnetic interaction}\\
{\huge at two loops}\\[5mm]
A.~Czarnecki\\
Institut f\"ur Theoretische Teilchenphysik, Universit\"at Karlsruhe,\\
D--76128 Karlsruhe, Germany\\
A.~G.~Grozin\\
Budker Institute of Nuclear Physics, Novosibirsk 630090, Russia
\end{center}

\begin{abstract}
We present the coefficient of the chromomagnetic interaction operator,
the only unknown coefficient in the Heavy Quark Effective Theory
(HQET) lagrangian up to the $1/m$ level,
with the two--loop accuracy by matching scattering amplitudes
of an on--shell heavy quark in an external field in full QCD and HQET,
and obtain the two--loop anomalous dimension of this operator in HQET.
\end{abstract}

\section{Introduction}
\label{Intro}

The leading order HQET lagrangian~\cite{EH1}
\begin{equation}
L_0 = \overline{Q}_v ivD Q_v
\label{L0}
\end{equation}
has a unit coefficient by construction
(here $Q_v=\slash{v}Q_v$ is a static quark field with velocity $v$).
At the $1/m$ level two new terms appear~\cite{EH2,FGL}
\begin{equation}
L = L_0
+ \frac{C_k(\mu)}{2m} \overline{Q}_v \left((vD)^2-D^2\right) Q_v
+ \frac{C_m(\mu)}{4m} \overline{Q}_v G_{\mu\nu} \sigma^{\mu\nu} Q_v
\label{L1}
\end{equation}
(where the composite operators are normalized at the scale $\mu$).
For the coefficient of the kinetic--energy operator
\begin{equation}
C_k(\mu) = 1
\label{Ck}
\end{equation}
holds in all orders of perturbation theory,
due to reparametrization invariance of HQET~\cite{LM,Ch,FGM}.
Only the coefficient $C_m(\mu)$ of the chromomagnetic interaction
operator is not known exactly.
It can be found by matching scattering amplitudes of an on--shell heavy quark
in an external chromomagnetic field in QCD and HQET up to $1/m$ terms.
This was done in~\cite{EH2} at the one--loop level;
the one--loop anomalous dimension of the chromomagnetic interaction
operator is therefore known~\cite{EH2,FGL}.
It is natural to perform matching at $\mu\approx m$,
where $C_m(\mu)$ contains no large logarithm.
Renormalization group can be used to obtain $C_m$ at $\mu\ll m$:
\begin{equation}
C_m(\mu) = C_m(m) \exp\left(-\int_{\alpha_s(m)}^{\alpha_s(\mu)}
\frac{\gamma_m(\alpha)}{2\beta(\alpha)} \frac{d\alpha}{\alpha} \right).
\label{RG}
\end{equation}
Two--loop anomalous dimension $\gamma_m=\frac{d\log Z_m}{d\log\mu}
=\gamma_1\frac{\alpha_s}{4\pi}+\gamma_2\left(\frac{\alpha_s}{4\pi}\right)^2
+\cdots$
and $\beta$--function $\beta=-\frac{1}{2}\frac{d\log\alpha_s}{d\log\mu}
=\beta_1\frac{\alpha_s}{4\pi}+\beta_2\left(\frac{\alpha_s}{4\pi}\right)^2
+\cdots$
should be used together with one--loop terms in
$C_m(m)=1+C_1\frac{\alpha_s(m)}{4\pi}+\cdots$~\cite{EH2}.
Chromomagnetic interaction is the only term violating the heavy--quark
spin symmetry~\cite{IW} at the $1/m$ level.
Numerous applications of the lagrangian~(\ref{L0}--\ref{L1})
are reviewed in~\cite{N}.

In this paper, we obtain $C_m(\mu)$ at two loops from QCD/HQET matching.
We consider scattering amplitude of an on--shell heavy quark
with the initial momentum $p_1=mv$ and a final momentum $p_2$
in a weak external field $A^a_\mu$ to the linear order in $q=p_2-p_1$.
Similarly to~\cite{EH2}, we use dimensional regularization
(in $d=4-2\varepsilon$ dimensions) and the background field
formalism~\cite{A}
in which the combination $gA^a_\mu$ is not renormalized.
We consider QCD with $n_l$ massless flavours and a single heavy flavour $Q$;
the effect of a massive flavour with a different mass
will be considered elsewhere.
There are two possible effective theories:
with loops of the heavy flavour $Q$ and without such loops.
Matching on--shell matrix elements produces finite results in both theories,
in contrast to the case of a heavy--light current~\cite{BG}.
The HQET diagrams contain no scale and hence vanish in dimensional
regularization,
except for those with a massive quark loop (Section~\ref{HQET}).
QCD on--shell diagrams were independently calculated
with the REDUCE~\cite{H,G} package RECURSOR~\cite{B2}
and a FORM~\cite{V} package~\cite{C} (Section~\ref{QCD}).
Comparing the HQET and QCD matrix elements, we obtain
the matching coefficient $C_m(m)$ and its anomalous dimension $\gamma$
(Section~\ref{Res}).

\section{HQET calculation}
\label{HQET}

The scattering matrix element of an on--shell quark in HQET up to $1/m$ level
has the structure
\begin{equation}
M = \overline{u}_v(q) gA^a_\mu t^a
\left( v^\mu + C_k Z_k^{-1} \tilde{\varepsilon} \frac{q^\mu}{2m}
+ C_m Z_m^{-1} \tilde{\mu} \frac{[\gamma^\mu,\slash{q}]}{4m} \right) u_v(0)\,,
\label{Me}
\end{equation}
where $qv=0$;
$\tilde{\varepsilon}=\tilde{Z}_Q\tilde{\varepsilon}_0$,
$\tilde{\mu}=\tilde{Z}_Q\tilde{\mu}_0$;
$\tilde{\varepsilon}_0$ and $\tilde{\mu}_0$
are the bare proper vertex functions of the unrenormalized operators
$\overline{Q}_v \left((vD)^2-D^2\right) Q_v$ and
$\overline{Q}_v G_{\mu\nu}\sigma^{\mu\nu} Q_v$.
In HQET with $n_l$ massless flavours in loops,
all loop corrections vanish because they contain no scale:
$\tilde{Z}_Q=1$, $\tilde{\varepsilon}_0=1$, $\tilde{\mu}_0=1$.

In HQET with the heavy flavour $Q$ in loops,
the on--shell wave--function renormalization constant was found in~\cite{BG}:
\begin{eqnarray}
&&\tilde{Z}_Q = 1 - C_F T_F \frac{g_0^4 m^{-4\varepsilon}}{(4\pi)^d} I_0^2
\frac{(d-1)(d-2)(d-6)}{2(d-5)(d-7)} \,,
\label{ZQ}\\
&&I_0 = \frac{1}{\pi^{d/2}} \int \frac{d^d k}{1-(v+k)^2}
= \frac{2}{d-2} \Gamma(\varepsilon)\,;
\label{I0}
\end{eqnarray}

\begin{figure}[ht]
\begin{center}
\includegraphics{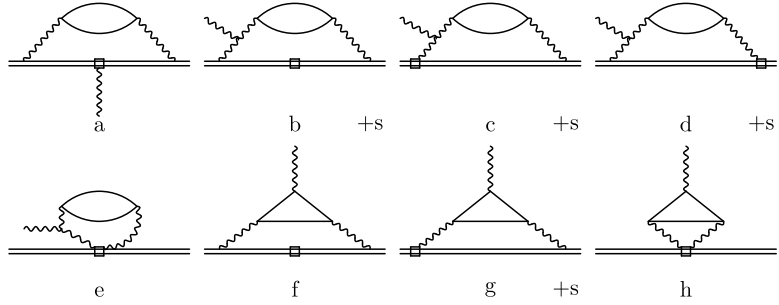}
\end{center}
\caption{Diagrams for $\tilde{\varepsilon}_0$;
${}+\mathrm{s}$ means adding the mirror--symmetric diagram.}
\label{F1}
\end{figure}

Diagrams for $\tilde{\varepsilon}_0$ are shown in Fig.~\ref{F1}.
We consider $A^a_\mu v^\mu=0$, and therefore diagrams in which
the background field is attached to the leading--order HQET vertex
$igt^a v^\mu$ do not contribute.
Diagrams in which the background field is attached to a four--leg $1/m$ vertex
do not depend on $q$ and do not contribute to the structure linear in $q$.
The diagrams Fig.~\ref{F1}e, h have zero colour factor.
A method of calculation of scalar integrals in such diagrams was proposed
in~\cite{BG}: we average over directions of $v$ in scalar integrals and obtain
\begin{equation}
\overline{(vk)^n} =
\frac{\Gamma\left(\frac{n+1}{2}\right)}{\Gamma\left(\frac{1}{2}\right)}
\frac{\Gamma\left(\frac{d}{2}\right)}{\Gamma\left(\frac{d+n}{2}\right)}
(k^2)^{n/2}
\label{av}
\end{equation}
for even $n$ (positive or negative) and $0$ for odd $n$.
After that, two--loop massive bubble integrals remain.
An explicit formula for them can be found in~\cite{GBGS}.
We perform calculations in an arbitrary covariant gauge,
and check that the sum of diagrams is gauge--invariant.
Only the diagram Fig.~\ref{F1}a contains the colour structure $C_F T_F$;
it is easy to see that this contribution is exactly compensated
by $\tilde{Z}_Q$~(\ref{ZQ}).
In order to calculate $\tilde{\varepsilon}_0$, we can contract the diagrams
with $q$ in the polarization index of the external gluon,
and extract the $q^2$ part.
In the backgroung field formalism, such external--gluon insertions
produce the difference of the propagators with the original momentum
and the momentum shifted by $q$ (after separating colour factors).
Let's consider the diagrams Fig.~\ref{F1}c, d, g first.
Most of the terms in the sum cancel each other,
leaving the difference of two terms:
the diagram without the external gluon,
and the same diagram with the heavy--quark momentum shifted by $q$.
This shift does not influence the propagator;
the kinetic energy vertex contains a term linear in $q$,
which vanishes after the loop integrations.
The sum of the diagrams Fig.~\ref{F1}b, f gives a similar difference
of two diagrams without the external gluon.
The difference of the two--leg kinetic energy vertices
now contains a $q^2$ term,
which is multiplied by the same integral as in~(\ref{ZQ}).
Finally, we obtain
\begin{equation}
\tilde{\varepsilon} = 1\,.
\label{ee}
\end{equation}
This fact is crucial for the proof of~(\ref{Ck}).
Note that the ``all--order'' proof of the 
reparametrization invariance~\cite{FGM}
ignores all massive loops in HQET (of the external heavy flavour or any
other massive quark) and therefore is valid up to the one--loop level
only. 

In the case of $\tilde{\mu}_0$, the diagrams Fig.~\ref{F1}b, f do not exist.
The colour factors of Fig.~\ref{F1}e, h no longer vanish.
Again, the only $C_F T_F$ contribution of Fig.~\ref{F1}a is compensated
by $\tilde{Z}_Q$ and we obtain
\begin{equation}
\tilde{\mu} = 1 + C_A T_F \frac{g_0^4 m^{-4\varepsilon}}{(4\pi)^d}
I_0^2 \frac{(d-2)(d^2-9d+16)}{4(d-5)(d-7)} \,.
\label{me}
\end{equation}

\section{QCD calculation}
\label{QCD}

The scattering matrix element of an on--shell quark in a weak external field
has the structure
\begin{equation}
M = \overline{u}(p_2) gA^a_\mu t^a
\left(\varepsilon(q^2)\frac{(p_1+p_2)^\mu}{2m}
+\mu(q^2)\frac{[\gamma^\mu,\slash{q}]}{4m}\right) u(p_1)\,.
\label{M}
\end{equation}
Up to the 
linear terms in $q$ it is determined by the total quark colour charge
$\varepsilon=Z_Q\varepsilon_0(0)=1$
and chromomagnetic moment $\mu=Z_Q\mu_0(0)$.
The on--shell wave--function renormalization $Z_Q$ was obtained in~\cite{BGS}.
Let us see why $\varepsilon=1$.
Diagrams for the bare proper vertex $\Lambda_0^\mu$
in the constant background field $A_\mu$ can be obtained from
the diagrams for the bare mass operator $\Sigma_0(p)$ by shifting $p$:
$\Sigma_0(p+A)=\Sigma_0(p)-\Lambda_0^\mu A_\mu$.
The term $\varepsilon_0$ is separated by the $\gamma$--matrix projector
$\varepsilon_0=1+\frac{1}{4}\Tr\Lambda_0^\mu v_\mu(\slash{v}+1)$.
On the mass shell,
$\frac{1}{4}\Tr(\Sigma_0(mv+A)-\Sigma_0(mv))(\slash{v}+1)=(1-Z_Q^{-1})vA$.
This gives $\varepsilon_0=Z_Q^{-1}$.
This argument is valid for both an abelian and a nonabelian background
field $A$. 

\begin{figure}[p]
\begin{center}
\includegraphics{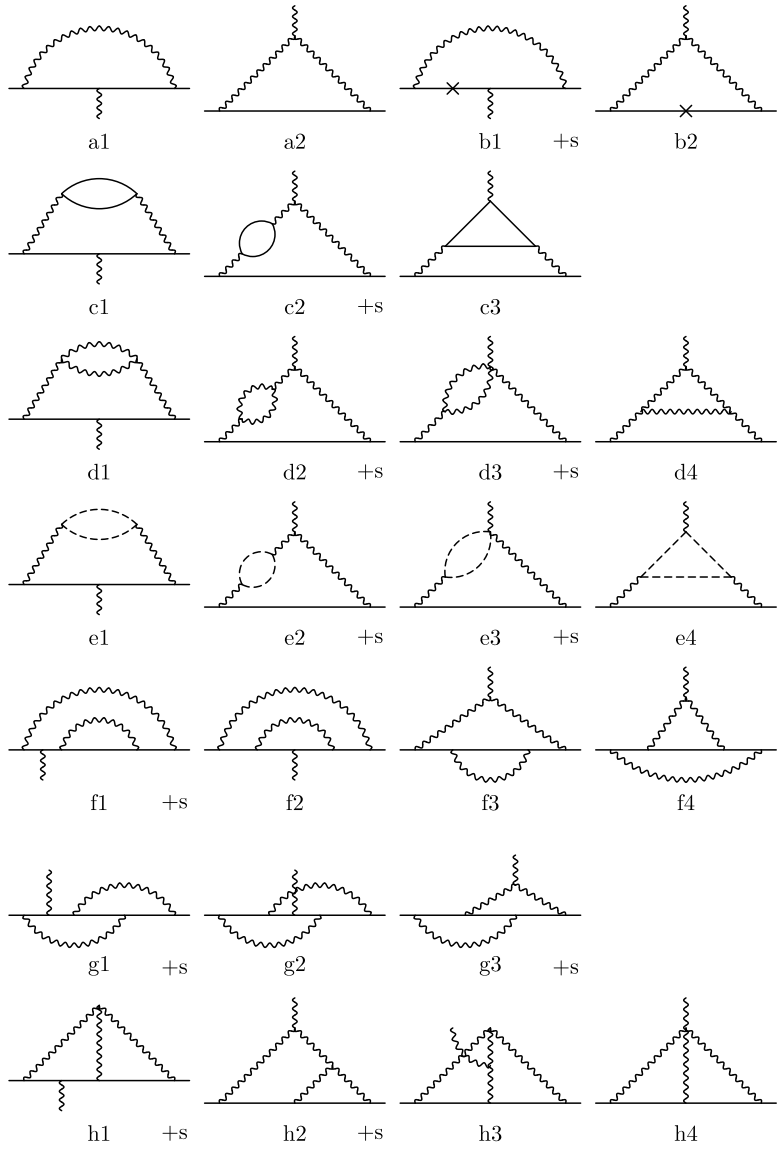}
\end{center}
\caption{Diagrams for the QCD proper vertex.}
\label{F2}
\end{figure}

We calculate bare proper vertices $\varepsilon_0$, $\mu_0$
on the renormalized mass shell.
Therefore, it is convenient to use the pole mass $m$ in the lagrangian,
and to incorporate the vertex
produced by the mass counterterm $\Delta m$~\cite{GBGS}.
Diagrams for the proper vertex can be obtained from those
for the mass operator by inserting the background field vertex
in all possible ways (Fig.~\ref{F2}).
Using integration by parts recurrence relations~\cite{GBGS,BGS},
all two--loop on--shell integrals can be reduced to $I_0^2$ and
\begin{eqnarray}
&&I_1 = \frac{1}{\pi^d} \int \frac{d^d k d^d l}{(-k^2)(-(l-k)^2)(1-(v+l)^2)}
\nonumber\\
&&\quad{} = \frac{4(2d-7)}{(d-3)(3d-8)(3d-10)}
\frac{\Gamma^2(1-\varepsilon)\Gamma(1+2\varepsilon)\Gamma(1-4\varepsilon)}
{\Gamma(1+\varepsilon)\Gamma(1-2\varepsilon)\Gamma(1-3\varepsilon)}
\Gamma^2(\varepsilon)\,,
\nonumber\\
&&I_2 = \frac{1}{\pi^d} \int \frac{d^d k d^d l}
{(1-(v+k)^2)(1-(v+l)^2)(1-(v+k+l)^2)}
\label{I12}\\
&&\quad{} = \frac{3(d-2)^2(5d-18)}{2(d-3)(3d-8)(3d-10)} I_0^2
- 2 \frac{d-4}{2d-7} I_1 - \frac{16(d-4)^2}{(3d-8)(3d-10)} I(\varepsilon)\,,
\nonumber
\end{eqnarray}
where~\cite{B1,B2}
\begin{equation}
I(\varepsilon) = I + O(\varepsilon), \quad
I = \pi^2 \log 2 - \frac{3}{2}\zeta(3).
\end{equation}
The diagrams Fig.~\ref{F2}c with massless quarks
(colour structures $C_F T_F n_l$ and $C_A T_F n_l$)
and Fig.~\ref{F2}d, e contain only $I_1$.
The integral $I_2$ is contained only in Fig.~\ref{F2}c with the heavy quark
(colour structures $C_F T_F$ and $C_A T_F$) and Fig.~\ref{F2}g.
The diagram Fig.~\ref{F2}h3 has zero colour factor.

We perform all calculations in an arbitrary covariant gauge,
and check that the $d$--dimensional results for $\varepsilon_0$ and $\mu_0$
are gauge--invariant.
We check that $\varepsilon_0=Z_Q^{-1}$; the same equality holds
if we use colour factors for an abelian background fields in $\varepsilon_0$.
Moreover, this is true for each group of diagrams obtained from
a single diagram for the mass operator: the sum of these diagrams,
with each set of colour factors, gives the contribution of the original
diagram to $Z_Q^{-1}$.
This provides a strong check of our procedures.
The programs for calculating $\mu_0$ are obtained by replacing
only the $\gamma$--matrix projector and hence are reliable.
If we use colour factors for an abelian background field, we reproduce
the heavy--quark magnetic moment~\cite{FT1};
if the dynamical gauge field is also abelian,
the classic result for the QED electron magnetic moment~\cite{SP} is reproduced
(in dimensional regularization, it was discussed in~\cite{FT2}).
In view of all these checks, we are confident in our final result
for the heavy--quark chromomagnetic moment.

The full $d$--dimensional result is presented in the Appendix.
Expanding it in $\varepsilon$ and re--expressing it via $\alpha_s(\mu)$,
we obtain
\begin{eqnarray}
\hspace{-7mm}&&\mu = 1 + \frac{\alpha_s(\mu)}{4\pi} e^{-2L\varepsilon}
\left(1-\frac{\alpha_s}{4\pi\varepsilon}\beta_1\right)
\left[2C_F(1+4\varepsilon) + C_A\left(\frac{1}{\varepsilon}+2
+\frac{\pi^2}{12}\varepsilon+2\varepsilon\right)\right]
\nonumber\\
\hspace{-7mm}&&\quad{} + \left(\frac{\alpha_s}{4\pi}\right)^2
e^{-4L\varepsilon}
\Biggl[ C_F^2 \left(-8I+\frac{20}{3}\pi^2-31\right)
+ C_F C_A \left(\frac{28}{3}\frac{1}{\varepsilon}
+\frac{4}{3}I+\frac{4}{3}\pi^2+\frac{605}{9}\right)
\nonumber\\
\hspace{-7mm}&&\quad{} + C_A^2 \left(\frac{7}{3}\frac{1}{\varepsilon^2}
+\frac{101}{9}\frac{1}{\varepsilon}+\frac{4}{3}I-\frac{3}{2}\pi^2
+\frac{1057}{27}\right)
\label{mu}\\
\hspace{-7mm}&&\quad{} + C_F T_F n_l \left(-\frac{8}{3}\frac{1}{\varepsilon}
-\frac{196}{9}\right)
+ C_A T_F n_l \left(-\frac{2}{3}\frac{1}{\varepsilon^2}
-\frac{37}{9}\frac{1}{\varepsilon}-\frac{5}{9}\pi^2-\frac{371}{27}\right)
\nonumber\\
\hspace{-7mm}&&\quad{} + C_F T_F \left(-\frac{8}{3}\frac{1}{\varepsilon}
-\frac{16}{3}\pi^2+\frac{380}{9}\right)
+ C_A T_F \left(-\frac{4}{3}\frac{1}{\varepsilon^2}
-\frac{8}{3}\frac{1}{\varepsilon}+\frac{8}{9}\pi^2-\frac{370}{27}\right)
\Biggr]\,,
\nonumber
\end{eqnarray}
where $L=\log m/\mu$, $\beta_1=\frac{11}{3}C_A-\frac{4}{3}T_F n_f$,
$n_f=n_l+1$.

\section{Results}
\label{Res}

The coefficients $C_k$, $C_m$ in the HQET lagrangian~(\ref{L1})
are tuned in such a way that the full QCD matrix element~(\ref{M}),
expanded to linear terms in $q$,
is equal to the HQET matrix element~(\ref{Me}).
Infrared (or on--shell) singularities are the same in both theories.
QCD and HQET spinors are related by~\cite{N} $u(p_1)=u_v(0)$,
$u(p_2)=\left(1+\frac{\slash{q}}{2m}\right)u_v(q)$;
the extra term with $\slash{q}$ should only be taken into account
in zeroth order terms, but then it vanishes.

We consider HQET with $n_l$ light flavours in loops first.
Comparing the structure $q^\mu$, we obtain $Z_k^{-1}C_k=1$,
and hence $C_k(\mu)=1$~(\ref{Ck}).
Comparing the structure $[\gamma^\mu,\slash{q}]$, we obtain
$Z_m^{-1}C_m=\mu/\tilde{\mu}$.
We find $Z_m$ from the requirement that $C_m$ is finite.
Terms $1/\varepsilon^2$ in it satisfy the consistency condition
which is necessary for the anomalous dimension to be finite
at $\varepsilon\to0$.
Terms $1/\varepsilon$ give the anomalous dimension
\begin{equation}
\gamma_m = 2 C_A \frac{\alpha_s}{4\pi}
+ \frac{4}{9} C_A (17 C_A - 13 T_F n_l) \left(\frac{\alpha_s}{4\pi}\right)^2\,.
\label{gam}
\end{equation}
The chromomagnetic interaction coefficient at $\mu=m$ is
\begin{eqnarray}
&&C_m(m) = 1 + 2(C_F+C_A) \frac{\alpha_s(m)}{4\pi}
+ \Biggl[ C_F^2 \left(-8I+\frac{20}{3}\pi^2-31\right)
\nonumber\\
&&\quad{} + C_F C_A \left(\frac{4}{3}I+\frac{4}{3}\pi^2+\frac{269}{9}\right)
+ C_A^2 \left(\frac{4}{3}I-\frac{17}{9}\pi^2+\frac{805}{27}\right)
\nonumber\\
&&\quad{} + C_F T_F n_l \left(-\frac{100}{9}\right)
+ C_A T_F n_l \left(-\frac{4}{9}\pi^2-\frac{299}{27}\right)
\label{Cm}\\
&&\quad{} + C_F T_F \left(-\frac{16}{3}\pi^2+\frac{476}{9}\right)
+ C_A T_F \left(\pi^2-\frac{298}{27}\right)
\Biggr] \left(\frac{\alpha_s}{4\pi}\right)^2\,.
\nonumber
\end{eqnarray}
With $N_c=3$ colours, its numerical value is
\begin{equation}
C_m(m) = 1 + \frac{16}{3} \frac{\alpha_s(m)}{\pi} +
\left( 21.79 - 1.91 n_l \right) \left(\frac{\alpha_s}{\pi}\right)^2\,.
\label{num}
\end{equation}
The two--loop correction is large.
The exact two--loop coefficient at $n_l=4$ is 40\% less than the expectation
based on the naive nonabelianization~\cite{BG},
i.~e.\ it is not particularly accurate,
but predicts the correct sign and order of magnitude.
The heavy--quark loop contributes merely $-0.10$
to the bracket in~(\ref{num}).

If we now include $Q$--loops in HQET, we still have $C_k(\mu)=1$~(\ref{Ck}).
The anomalous dimension~(\ref{gam}) now contains $n_f=n_l+1$ instead of $n_l$.
The chromomagnetic interaction coefficient~(\ref{Cm}) has the coefficient
of $C_A T_F$ equal to $\frac{10}{9}\pi^2-\frac{227}{27}$,
leading to $22.14$ in the bracket in~(\ref{num}).

Our main results are the anomalous dimension~(\ref{gam})
and the chromomagnetic interaction coefficient at $\mu=m$~(\ref{Cm}).
If $L=\log m/\mu$ is not very large, the best approximation to $C_m(\mu)$
is the exact two--loop matching formula
\begin{equation}
C_m(\mu) = 1 + \left(C_1 - \gamma_1 L \right) \frac{\alpha_s(m)}{4\pi}
+ \left[C_2 - \left(C_1\gamma_1+\gamma_2\right) L
+ \gamma_1\left(\gamma_1-\beta_1\right) L^2 \right]
\left(\frac{\alpha_s}{4\pi}\right)^2\,,
\label{Cmu}
\end{equation}
in which all terms
$(\alpha_s/\pi)^2 L^{2,1,0}$ are taken into account,
but $(\alpha_s/\pi)^3 L^3$ and other higher order terms are dropped.
Otherwise, it is better to sum leading and subleading logarithms
using~(\ref{RG}):
\begin{equation}
C_m(\mu)
= \left(\frac{\alpha_s(\mu)}{\alpha_s(m)}\right)^{-\gamma_1/(2\beta_1)}
\left[ 1 + C_1 \frac{\alpha_s(m)}{4\pi}
- \frac{\beta_1\gamma_2-\beta_2\gamma_1}{2\beta_1^2}
\frac{\alpha_s(\mu)-\alpha_s(m)}{4\pi}
\right]\,.
\label{CRG}
\end{equation}
These results can be applied to all cases of the spin symmetry
violation, such as $D$--$D^*$ and $B$--$B^*$ splittings,
$1/m$ corrections in $B\to D$ and $B\to D^*$ semileptonic decays,
etc.~\cite{N}.

In the course of this work, we were informed by M.~Neubert
about the ongoing calculation of the two--loop anomalous dimension
of the chromomagnetic interaction operator by a completely
different method~\cite{ABN}.
Our result~(\ref{gam}) agrees with~\cite{ABN}.

\textbf{Acknowledgements}.
We are grateful to M.~Neubert for communicating the result of~\cite{ABN}
before publication,
and to D.~J.~Broadhurst for numerous fruitful discussions
of HQET and methods of multiloop calculations.
A.~C.'s research was supported by the grant BMBF 057KA92P.

\newpage
\appendix

\section{Heavy--quark chromomagnetic moment}
\label{App}

The $d$--dimensional chromomagnetic moment has the form
\begin{eqnarray}
&&\mu = 1 + \frac{g_0^2 m^{-2\varepsilon}}{(4\pi)^{d/2}} \frac{1}{4}
\frac{d-2}{d-3} \left[ 2(d-4)(d-5)C_F - (d^2-8d+14)C_A \right] I_0
\nonumber\\
&&\quad{}+\frac{g_0^4 m^{-4\varepsilon}}{(4\pi)^d} \sum_{i,j} a_{ij}C_i J_j\,,
\nonumber
\end{eqnarray}
where the colour structures are $C_1=C_F^2$, $C_2=C_F C_A$, $C_3=C_A^2$,
$C_4=C_F T_F n_l$, $C_5=C_A T_F n_l$, $C_6=C_F T_F$, $C_7=C_A T_F$ and
\begin{eqnarray}
&&J_0 = \frac{(d-2)I_0^2}{16(d-1)(d-3)^2(d-4)^2(d-5)(d-6)(d-7)}\,,
\nonumber\\
&&J_1 = \frac{I_1}{16(d-1)(d-3)(d-4)(2d-7)}\,,
\quad
J_2 = \frac{I_2}{16(d-1)(d-4)^2(d-6)}
\nonumber
\end{eqnarray}
are chosen in such a way that the coefficients $a_{ij}$ are polynomial in $d$.
\LaTeX{} source of the following equation, presenting all nonvanishing
coefficients $a_{ij}$, was generated by a REDUCE program using the 
package RLFI by R.~Liska (see~\cite{H,G}):
\begin{eqnarray}
\hspace{-7mm}&&a_{11} =
4(d-5)(d-6)(d-7)(d^{8}-32 d^{7}+460 d^{6}-3828 d^{5}+19940 d^{4}-66032 d^{3}
\nonumber\\
\hspace{-7mm}&&\quad{}
+135065 d^{2}-155640 d+77356)
\nonumber\\
\hspace{-7mm}&&a_{12} =
-16(2 d-7)(2 d^{6}-53 d^{5}+557 d^{4}-3005 d^{3}+8828 d^{2}-13450 d+8336)
\nonumber\\
\hspace{-7mm}&&a_{13} =
8(d-6)(2 d^{6}-60 d^{5}+682 d^{4}-3871 d^{3}+11723 d^{2}-18066 d+11120)
\nonumber\\
\hspace{-7mm}&&a_{21} =
-2(d-6)(d-7)(2 d^{9}-72 d^{8}+1175 d^{7}-11329 d^{6}+70628 d^{5}-293546 d^{4}
\nonumber\\
\hspace{-7mm}&&\quad{}
+809949 d^{3}-1426799 d^{2}+1454100 d-652948)
\nonumber\\
\hspace{-7mm}&&a_{22} =
4(16 d^{7}-474 d^{6}+5777 d^{5}-37896 d^{4}+145361 d^{3}-327378 d^{2}+402044 d
\nonumber\\
\hspace{-7mm}&&\quad{}
-208160)
\nonumber\\
\hspace{-7mm}&&a_{23} =
-4(d-6)(5 d^{6}-146 d^{5}+1625 d^{4}-9065 d^{3}+27085 d^{2}-41398 d+25424)
\nonumber\\
\hspace{-7mm}&&a_{31} =
(d-6)(d-7)(d^{9}-35 d^{8}+557 d^{7}-5259 d^{6}+32250 d^{5}-132396 d^{4}
\nonumber\\
\hspace{-7mm}&&\quad{}
+362076 d^{3}-633794 d^{2}+642768 d-287288)
\nonumber\\
\hspace{-7mm}&&a_{32} =
-(3 d-8)(6 d^{6}-159 d^{5}+1672 d^{4}-9015 d^{3}+26460 d^{2}-40276 d+24928)
\nonumber\\
\hspace{-7mm}&&a_{33} =
2(d-3)(d-6)(3 d-8)(d^{4}-23 d^{3}+176 d^{2}-550 d+596)
\nonumber\\
\hspace{-7mm}&&a_{42} =
-64(d-2)(d-3)(d-4)^{2}(d^{2}-7 d+11)
\nonumber\\
\hspace{-7mm}&&a_{52} =
4(d-3)(d-4)(3 d-8)(2 d^{3}-19 d^{2}+57 d-56)
\nonumber\\
\hspace{-7mm}&&a_{61} =
4(d-2)(d-3)(d-4)(d-7)(3 d^{5}-85 d^{4}+875 d^{3}-4123 d^{2}+8898 d
\nonumber\\
\hspace{-7mm}&&\quad{}
-7008)
\nonumber\\
\hspace{-7mm}&&a_{63} =
-8(d-4)(d^{2}-9 d+16)(d^{3}-13 d^{2}+26 d+16)
\nonumber\\
\hspace{-7mm}&&a_{71} =
-(d-3)(d-4)(9 d^{7}-290 d^{6}+3822 d^{5}-26680 d^{4}+106477 d^{3}
\nonumber\\
\hspace{-7mm}&&\quad{}
-242974 d^{2}+294012 d-145896)
\nonumber\\
\hspace{-7mm}&&a_{73} =
2(d-4)(3 d-8)(d^{4}-16 d^{3}+75 d^{2}-96 d-28)
\nonumber
\end{eqnarray}

\end{document}